\documentclass[reprint,superscriptaddress,prb,longbibliography]{revtex4-2}
\usepackage[utf8]{inputenc}
\usepackage[T1]{fontenc}
\usepackage{amsmath}
\usepackage{amssymb}
\usepackage{mathrsfs}
\usepackage{graphicx}
\usepackage[capitalize, nameinlink]{cleveref}
\usepackage{subfigure}
\graphicspath{{figures/}}
\usepackage{adjustbox}
\usepackage{booktabs}
\usepackage{multirow}

\usepackage{color}
\usepackage{rotating}
\usepackage{ulem}
\usepackage{bm}
\begin{document}

\title{Dynamic annihilation pathways of magnetic skyrmions}

\author{Matthew Copus}
\affiliation{Center for Magnetism and Magnetic Nanostructures, University of Colorado Colorado Springs, Colorado Springs, CO 80918, USA}
\author{Ezio Iacocca}
\affiliation{Center for Magnetism and Magnetic Nanostructures, University of Colorado Colorado Springs, Colorado Springs, CO 80918, USA}

\date{\today}

\begin{abstract}
The investigation of magnetic solitons often relies on numerical modeling to determine key features such as stability, annihilation, nucleation, and motion. However, as soliton sizes approach atomic length scales, the accuracy of these predictions becomes increasingly sensitive to the details of the numerical model. Here, we study the annihilation of two-dimensional magnetic skyrmions using a pseudospectral approach and compare its performance to that of conventional micromagnetic simulations. A central distinction between the models lies in their treatment of the exchange interaction, which governs the magnon dispersion relation and plays a crucial role in balancing the uniaxial anisotropy to stabilize skyrmions. We demonstrate that both the choice of model and the spatial discretization significantly influence skyrmion dynamics and the magnetic field required for annihilation. The pseudospectral model provides a consistent description across length scales and captures complex behaviors such as skyrmion breathing on the path toward annihilation. Our results have direct implications for the state-of-the-art modeling of skyrmions and other two-dimensional textures and will also impact the modeling of three-dimensional textures such as hopfions.% More broadly, our approach will contribute to the development seamless multiscale model and optimization machine learning approaches for material discovery.
\end{abstract}

\maketitle

\section{Introduction}

Nanoscale magnetic textures continue to attract intense research interest due to their rich underlying physics~\cite{Kosevich1990,Braun2012} and promising technological applications~\cite{Fert2013,Sampaio2013,Tomasello2014,Jiang2015,Zhou2015,Gobel2021,Marrows2021}. Even though many possible magnetic textures have been theoretically known for decades~\cite{Kosevich1990}, only recent advances in fabrication and imaging techniques~\cite{Donnelly2015,Donnelly2017,Donnelly2021,Raftrey2021,Rana2023,DiPietro2023,Girardi2024} have demonstrated their existence and opened pathways for their control. Topological textures~\cite{Braun2012} are of particular interest because their existence is protected by an energy barrier. Much effort has been devoted to magnetic skyrmions~\cite{Fert2013,Nagaosa2013,Fert2017}, which are topologically protected in two dimensions (2D), and, more recently, to magnetic hopfions~\cite{Wang2019,Liu2020,Kent2021,Balakrishnan2023,Zhang2023,Zheng2023}, which are topologically protected in three dimensions (3D). Despite their topological protection, it is challenging to preserve the stability of such textures. In the case of skyrmions, their topological protection is strictly in 2D~\cite{Braun2012}, meaning that their stability in thin films is compromised by the spontaneous nucleation of hedgehog-antihedgehog pairs, or Bloch points~\cite{Tatara2014,Fujishiro2019}, which quickly annihilate at physical edges~\cite{Im2019,Yu2020}. In the case of hopfions, proximity to physical surfaces can lead to their annihilation, although several experiments~\cite{Kent2021,Zheng2023} and analytical studies~\cite{Rybakov2022} have provided evidence for their stabilization.

As material dimensions continue to shrink and topological textures approach atomic scales, the discrete nature of solid-state matter can no longer be neglected. Most theoretical investigations of topological textures rely on a continuum model of the magnetic exchange interaction~\cite{Kosevich1990,Rohart2016,Buttner2018,Bogdanov2020}. However, this approach introduces assumptions that break down at the atomic scale, e.g., a singular magnetization vector at a Bloch point. One solution to this problem is to employ discrete numerical models such as atomistic spin dynamics (ASD)~\cite{Eriksson2017,Evans2014}. Such models provide a means to accurately describe Bloch points~\cite{Andreas2014} and skyrmion annihilation~\cite{Bessarab2015,Bessarab2018}. However, ASD simulations are constrained to small system sizes due to computational resource limitations. Another solution is to use micromagnetic simulations~\cite{Brown1963b,Vansteenkiste1,Donahue2015,Heil2019,austrup1}. This method models the exchange interaction using a continuum approximation, and therefore is strictly valid only for small wavevectors or large cell sizes. However, because the model is implemented numerically on a spatial grid of discrete cells, accurate solutions are typically achieved with small cell dimensions. For this reason, micromagnetic simulations are assumed to be valid for cell dimensions smaller than the exchange length but may nevertheless become unphysical as cell dimensions approach the atomic lattice constant.

ASD and micromagnetic simulations can complement each other, thus multiscale approaches have been sought~\cite{Andreas2014,Poluektov2018}. For example, smooth skyrmion motion was achieved in a lattice transitioning from an atomic to a continuum description~\cite{Mendez2020}. Despite such successes, multiscale approaches involve a transition between different models, which requires knowledge of the underlying geometry and a sufficiently large computational domain. Alternatively, one could use a pseudospectral Landau-Lifshitz (PS-LL) description of magnetization dynamics, which models atomic and continuum dynamics on equal footing~\cite{rockwell}. The PS-LL model leverages the fact that the magnon dispersion relation captures the relevant physical interactions at every spatial and temporal scale, so that the transition to larger cells is seamless in principle. This model has so far proven successful in describing ultrafast remagnetization induced by transient gratings~\cite{rockwell,Foglia2024}. Here, we extend the PS-LL to include the antisymmetric exchange interaction derived from first principles and apply it to investigate the annihilation pathways of magnetic skyrmions.

In this work, we demonstrate that the choice of numerical model and spatial discretization can lead to quantitatively different predictions of skyrmion dynamics. Using the PS-LL model, implemented with two distinct exchange interaction kernels, alongside results from the widely used micromagnetic simulation package MuMax3~\cite{Vansteenkiste1}, we examine the contraction and annihilation of a Néel-type skyrmion in a 2D magnetic thin film with uniaxial anisotropy and interfacial Dzyaloshinskii–Moriya interaction (DMI). The dynamics are driven by an increasing out-of-plane magnetic field. While all models show similar behavior initially, striking differences emerge as the skyrmion radius approaches the exchange length. The PS-LL model accurately describes the dynamics at both the atomic and micromagnetic limits. However, micromagnetic models either overestimate or underestimate the exchange energy in the system, leading to an incorrect prediction of the skyrmion's topological protection and thus its eventual annihilation. Our work demonstrates that, even within the micromagnetic regime, a pseudospectral approach provides a much more accurate description of the magnetization dynamics because the exchange interaction is described appropriately up to the first Brillouin zone. Such accuracy is essential for modeling the nucleation, stability, and annihilation of skyrmions as well as their manipulation by external stimuli. From a much broader point of view, our investigation highlights the limitations of standard solvers and computational tools that must be carefully considered as physical systems approach their ultimate scales and dynamics escape the standard models and approximations. Dispersion relations can be typically obtained by first principles and experiments, e.g., neutron scattering, for a variety of physical systems, and from there can be directly implemented in pseudospectral approaches.

\section{Two-dimensional PS-LL model}

The PS-LL model is based on the Landau-Lifshitz equation given by
\begin{equation}
\label{eq:LL}
\frac{\partial}{\partial t}\mathbf{m} = -\mathbf{m}\times\mathbf{H}_\mathrm{eff} - \alpha\mathbf{m}\times\left(\mathbf{m}\times\mathbf{H}_\mathrm{eff}\right),
\end{equation}
where $\mathbf{m}$ denotes the magnetization vector normalized by the saturation magnetization, $M_s$, $\alpha$ is the Gilbert damping constant which can be used in the Landau-Lifshitz form provided $\alpha\ll1$, and the effective field $\mathbf{H}_\mathrm{eff}$ is given by
\begin{equation}
    \label{eq:Heff}
    \mathbf{H}_\mathrm{eff} = \gamma \mu_0 \left[\mathbf{H}_l - \mathcal{F}^{-1}\left\{\kappa(\mathbf{k})\mathbf{\hat{m}}\right\}\right],
\end{equation}
where $\gamma$ is the gyromagnetic ratio and $\mu_0$ is the vacuum permeability. This equation involves local and nonlocal contributions. The local contribution
\begin{equation}
    \label{eq:Hlocal}
    \mathbf{H}_l = H_0\hat{z} +H_km_z\hat{z}
\end{equation}
consists of an external magnetic field of magnitude $H_0$ and a uniaxial anisotropy field of magnitude $H_k$. The nonlocal contributions are incorporated in Fourier space through the term $\kappa(\mathbf{k})\mathbf{\hat{m}}$, where $\kappa(\mathbf{k})$ is a normalized nonlocal interaction kernel including exchange and Dzyaloshinskii-Moriya interaction (DMI), $\mathbf{\hat{m}}$ is the Fourier transform of the magnetization, and $\mathcal{F}^{-1}$ represents the inverse Fourier transform.

For the exchange interaction in 2D, the contribution to the kernel is taken as the dispersion relation scaled by the saturation magnetization~\cite{Foglia2024},
\begin{equation}
    \label{eq:exch_k}
\omega(\mathbf{k}) = -2M_s\left(\frac{\lambda_{ex}}{a}\right)^2\left[2 - \cos{(k_x a)} - \cos{(k_y a)}\right],
\end{equation}
where $\lambda_{ex}$ is the exchange length and $a$ is the lattice constant. This dispersion relation can be obtained from first principles by considering the nearest-neighbor interaction between localized moments, i.e., under the assumption of a Heisenberg exchange model~\cite{White2007}. Higher-order exchange terms could be incorporated to represent next-nearest-neighbor interactions~\cite{Rybakov2022}, but we restrict our current investigation to nearest-neighbor interactions only. Under this assumption, Eq.~\eqref{eq:exch_k} provides an accurate representation of the spin dynamics across the entire first Brillouin zone (FBZ). The dispersion relation reduces to the well-known micromagnetic, long-wavelength limit when $\mathbf{k}\rightarrow0$
\begin{equation}
    \label{eq:exch_k2}
\omega_\mathrm{\mu}(\mathbf{k}) = \lim_{\mathbf{k}\rightarrow0}\omega(\mathbf{k}) = -M_s\lambda_{ex}^2\left(k_x^2 + k_y^2\right).
\end{equation}
Therefore, the PS-LL model allows for a smooth transition from the atomistic to the continuum limits, based solely on the numerical discretization~\cite{rockwell}.

The DMI is considered to be interfacial, with a symmetry-breaking direction along the out-of-plane $\hat{\mathbf{z}}$ axis. Its contribution to the dispersion relation can be derived from the DMI energy, considering a square lattice, as detailed in Appendix~\ref{sec:DMI}. Because DMI is asymmetric, it is convenient to represent the kernel as field components for each spatial direction independently:
\begin{subequations}
\label{eq:DMI}
\begin{eqnarray}
\label{eq:DMIx}
    \mathbf{H}_\mathrm{DMI}\cdot\hat{\mathbf{x}} &=& \mathcal{F}^{-1}\left\{-\frac{i2D}{\mu_0 M_s a}\sin{(k_x a)}\;\hat{m}_z\right\},\\
\label{eq:DMIy}
    \mathbf{H}_\mathrm{DMI}\cdot\hat{\mathbf{y}} &=& \mathcal{F}^{-1}\left\{-\frac{i2D}{\mu_0 M_s a}\sin{(k_y a)}\;\hat{m}_z\right\},\\
\label{eq:DMIz}
    \mathbf{H}_\mathrm{DMI}\cdot\hat{\mathbf{z}} &=& \mathcal{F}^{-1}\Big\{\frac{i2D}{\mu_0 M_s a}[\sin{(k_x a)}\;\hat{m}_y\nonumber\\&&\quad\quad\quad\quad\quad + \sin{(k_y a)}\;\hat{m}_x]\Big\}.
\end{eqnarray}
\end{subequations}
where $D$ is the interfacial DMI constant, with units of J~m$^{-2}$.

It is important to note that the DMI kernel is complex, which corresponds to a phase shift in Fourier space. Additionally, the presence of sine terms indicates that the kernel is an odd function, a property that, together with the phase shift, is the origin of nonreciprocity in the PS-LL description. Although the kernel is complex, its inverse Fourier transform is real-valued, as expected from a physical field. In the long-wavelength limit ($\mathbf{k}\rightarrow0$), a Taylor expansion of the sine functions shows that the kernel becomes linear in $k$, and the lattice constant vanishes. This is the regime where DMI is traditionally considered to be important as it produces a wavenumber shift in the dispersion relation~\cite{Nembach2015}. However, as wavelengths become shorter, the DMI scaling in Eqs~\eqref{eq:DMI} can increase by a factor $1/a\approx2.5$.
\begin{figure*}[t]
\includegraphics[width=\linewidth]{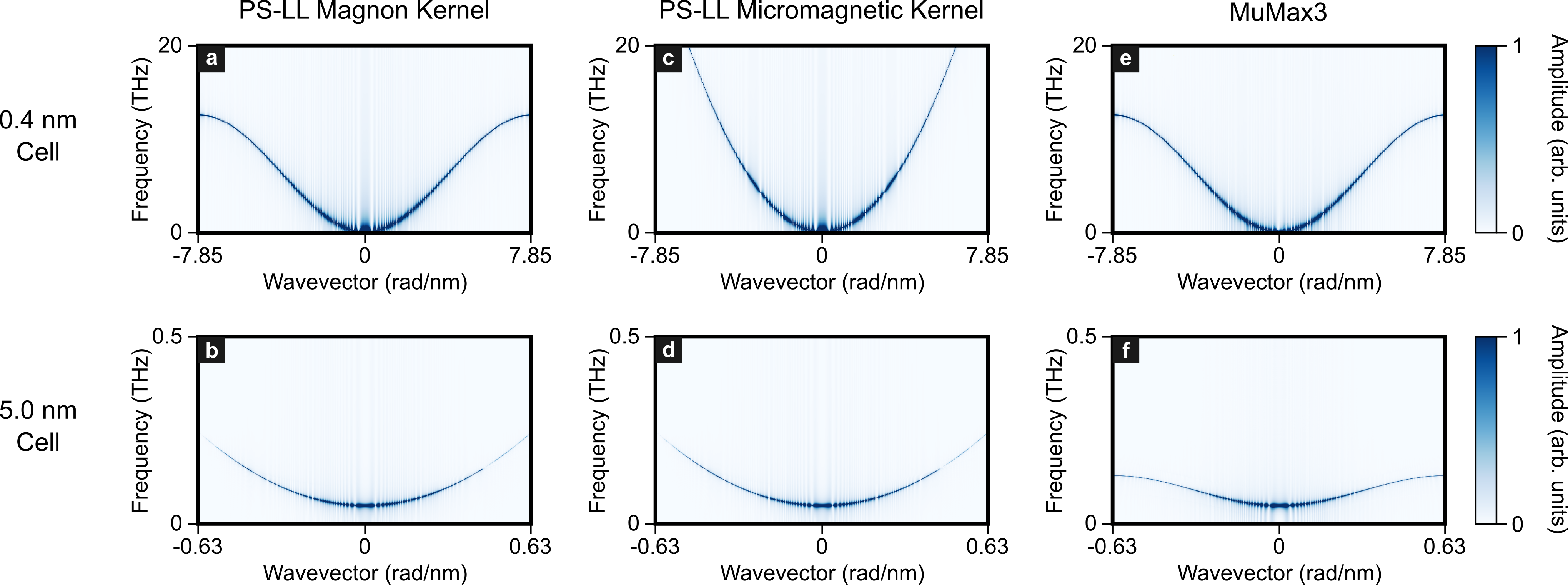}
\caption{\label{fig:dispersions} Magnon dispersion relations extracted from dynamic simulations using three different models: the PS-LL model with the full magnon kernel, (a) and (b), the PS-LL model with the micromagnetic kernel, (c) and (d), and MuMax3 (e) and (f). The top row shows results at an atomic-scale discretization of 0.4~nm, while the bottom row uses a micromagnetic-scale discretization of 5.0~nm. (a) The magnon kernel at an atomic resolution reproduces the cosine-shaped dispersion relation for the first Brillouin zone. (b) The magnon kernel at 5.0~nm cell size produces a dispersion curve in a confined region [12.5 times smaller than panel (a)] of the first Brillouin zone, revealing a parabolic form consistent with the low-$k$ limit. (c) The micromagnetic kernel an atomic resolution produces a parabolic dispersion which approximates this dispersion well near the origin but fails to capture short-wavelength behavior. (d) The micromagnetic kernel at 5.0~nm produces a dispersion curve that shows excellent agreement with panel (b), confirming the expected parabolic behavior at this scale. (e) MuMax3 at an atomic resolution closely matches the magnon kernel at atomic resolution due to its nearest-neighbor finite-difference scheme. (f) MuMax3 at 5.0~nm produces a cosine-shaped dispersion due to the use of a finite-difference scheme and therefore deviates from the other two models at this scale. The color scheme is found in Ref.~\cite{colorscheme}. }
\end{figure*}

The two energy contributions required to stabilize solitons~\cite{Kosevich1990} are contained within the effective field $\mathbf{H}_\mathrm{eff}$ from Eq.~\eqref{eq:Heff}. The uniaxial anisotropy, contained in the local field term $\mathbf{h}_l$ introduces nonlinearities, while the nonlocal field term $\mathcal{F}^{-1}\left\{\kappa(\mathbf{k})\mathbf{\hat{m}}\right\}$ accounts for dispersion. Most theoretical understanding of magnetic solitons can be obtained within the micromagnetic limit~\cite{Kosevich1990,Bogdanov2020}. However, exact solutions are typically nonlinear, so their dynamics and decay can generally only be treated approximately~\cite{Heil2019,austrup1}. Numerical methods are therefore essential for accurately describing soliton behavior in its entirety. For this reason, it is critical to assess the limitations that arise when implementing a specific numerical model. In particular, since the magnon dispersion relation quantifies the energy of non-collinear spin configurations, a consistent dispersion relation across spatial scales is required for correct soliton modeling. It is also important to note that the dispersion relation is derived under the assumption of small-amplitude, linear waves, but its physical origin does not rely on linearity.  This is the key advantage of the pseudospectral model: the dispersion relation serves as a kernel that quantifies the Heisenberg interaction and can, in fact, be derived directly from first principles as shown above for the DMI and as traditionally done to obtain the magnon dispersion relation~\cite{White2007}.

To demonstrate grid independence and the effect of numerical approximations, Fig.~\ref{fig:dispersions} shows the dispersion relations extracted from simulations using the PS-LL model with the magnon kernel $\omega(\mathbf{k})$, the PS-LL model with the micromagnetic kernel $\omega_\mathrm{\mu}(\mathbf{k})$, and MuMax3~\cite{Vansteenkiste1}, which is an open-source GPU-accelerated software that calculates the exchange interaction based on a finite difference method. For these simulations, we use material parameters consistent with Pt/Co/X multilayers, where X represents various metallic materials, $M_s = 1120$~kA~m$^{-1}$, exchange stiffness $A = 10.0$~pJ~m$^{-1}$, uniaxial anisotropy field $H_k = 1370$~kA~m$^{-1}$, and DMI constant $D = 1.50$~mJ~m$^{-2}$~\cite{Cheng2023PtCoCuSkyrmions,Lin2018PtCoWskyrmions}. For these material parameters, the exchange length is $\lambda_\mathrm{ex}=\sqrt{2A/(\mu_0M_s^2)}=3.56$~nm.

The dispersion relation is obtained from the dynamic evolution of a Gaussian initial condition set with a standard deviation equal to the cell size, effectively exciting the full available spectrum due to its broadband nature. By recording the decay of this initial condition, the magnon modes across the entire first Brillouin zone are populated. A fast Fourier transform is then applied to the resulting time-dependent $m_z$ magnetization component in both space and time from which a two-dimensional cross-section yields the dispersion relations shown in Fig.~\ref{fig:dispersions}. A Gilbert damping of $\alpha=0.005$ is used to maximize the contrast in Fourier space.

Because the material has perpendicular magnetic anisotropy (PMA), the Gaussian initial condition was set as a deviation from $m_z=1$, with amplitude 0.9, such that the minimum $m_z=0.1$ occurred at the Gaussian's center. The condition $|\mathbf{m}|=1$ was enforced by setting $m_x=\sqrt{1-m_z^2}$ and $m_y=0$ as the initial condition. We note that the Gaussian was set with an amplitude smaller than 1 to avoid the onset of nonlinear effects that would lead to the mixing of spectral components. Therefore, the dispersion relation can be reproduced for any Gaussian amplitude under 1.

To isolate the effect of the dispersion kernel, the demagnetizing field was approximated by setting a local contribution with magnitude equal to the saturation magnetization, $M_\mathrm{s}$. In other words, the anisotropy field $H_k\rightarrow H_k-M_s$. This approximation captures the dominant out-of-plane demagnetization effects in thin films, which is the regime considered here.

The obtained dispersion relations with the PS-LL magnon kernel are shown in Figs.~\ref{fig:dispersions}(a) and ~\ref{fig:dispersions}(b) for atomic and 5.0~nm cell sizes, respectively. For the atomic discretization of $0.4$~nm, we used a domain of $204.8$~nm~$\times~204.8$~nm, yielding a wavevector resolution of $0.031$~rad~nm$^{-1}$. The simulation was evolved for $10$~ps and sampled at $10$~fs intervals, resulting in a frequency resolution of $100$~GHz and a Nyquist frequency of $50$~THz. The $5.0$~nm cell size is longer than exchange length. In this case, we used a domain of $2560$~nm~$\times~2560$~nm, yielding a wavevector resolution of $2.5$~rad~$\mu$m$^{-1}$. The simulation was evolved for $1.0$~ns and sampled at $1.0$~ps intervals, resulting in a frequency resolution of $1$~GHz and a Nyquist frequency of $0.5$~THz. In both cases, the extracted dispersion relation matches the theoretical kernel across all wavevectors. Using a $5.0$~nm cell-size returns a dispersion that closely follows a $k^2$ dependence, in agreement with Eq.~\eqref{eq:exch_k2}. This behavior is expected near $k=0$, where the magnon dispersion is well approximated by the leading term of its Taylor expansion. For comparison, Figs.~\ref{fig:dispersions}(c) and~\ref{fig:dispersions}(d) show the dispersion relations obtained using the micromagnetic kernel for atomic and $5.0$~nm cell sizes, respectively. In both cases, the result is purely parabolic, as expected. The close agreement between Fig.~\ref{fig:dispersions}(c) and Fig.~\ref{fig:dispersions}(d) further confirms the grid independence of the PS-LL model using the magnon kernel, even for cell sizes above the exchange length.

We now compare these results with those obtained using MuMax3. Interestingly, for the atomic-scale discretization shown in Fig.~\ref{fig:dispersions}(e), the dispersion closely matches that of the PS-LL model with the magnon kernel. This agreement arises from MuMax3’s finite-difference implementation of the exchange interaction, which uses a nearest-neighbor approximation. Such an approximation coincides with the Taylor expansion of a cosine function. This connection becomes even more apparent at the $5.0$~nm cell-size, shown in Fig.~\ref{fig:dispersions}(f), where the resulting dispersion also exhibits a cosine functional form. It is important to stress that even though the cell-size is larger than the exchange length, the simulation is well-behaved and numerical instabilities are not observed at any time. From these results, we conclude that the numerically computed dispersion relation depends on the manner in which the exchange interaction is implemented: whether through increasing the finite-difference stencil, as in OOMMF~\cite{Donahue2015}, including higher-order terms in the dispersion expansion~\cite{Heil2019}, or adopting a finite-element approach.

{As a consequence of the linearization of the Landau-Lifshitz equation, the dispersion relation is often overlooked when considering nonlinear structures such as solitons. However, the dispersion relation encodes the coupling between space and time in the system and thus fully describes the interactions between magnetic moments, regardless of the system’s linearity. As we demonstrate below, the discrepancies between the dispersion relations from different models} become relevant when examining the annihilation of topological textures such as skyrmions. Likewise, these discrepancies could also impact the nucleation and stability of such textures.

\section{Dynamic skyrmion annihilation}

\begin{figure}
\includegraphics[width=3.3in]{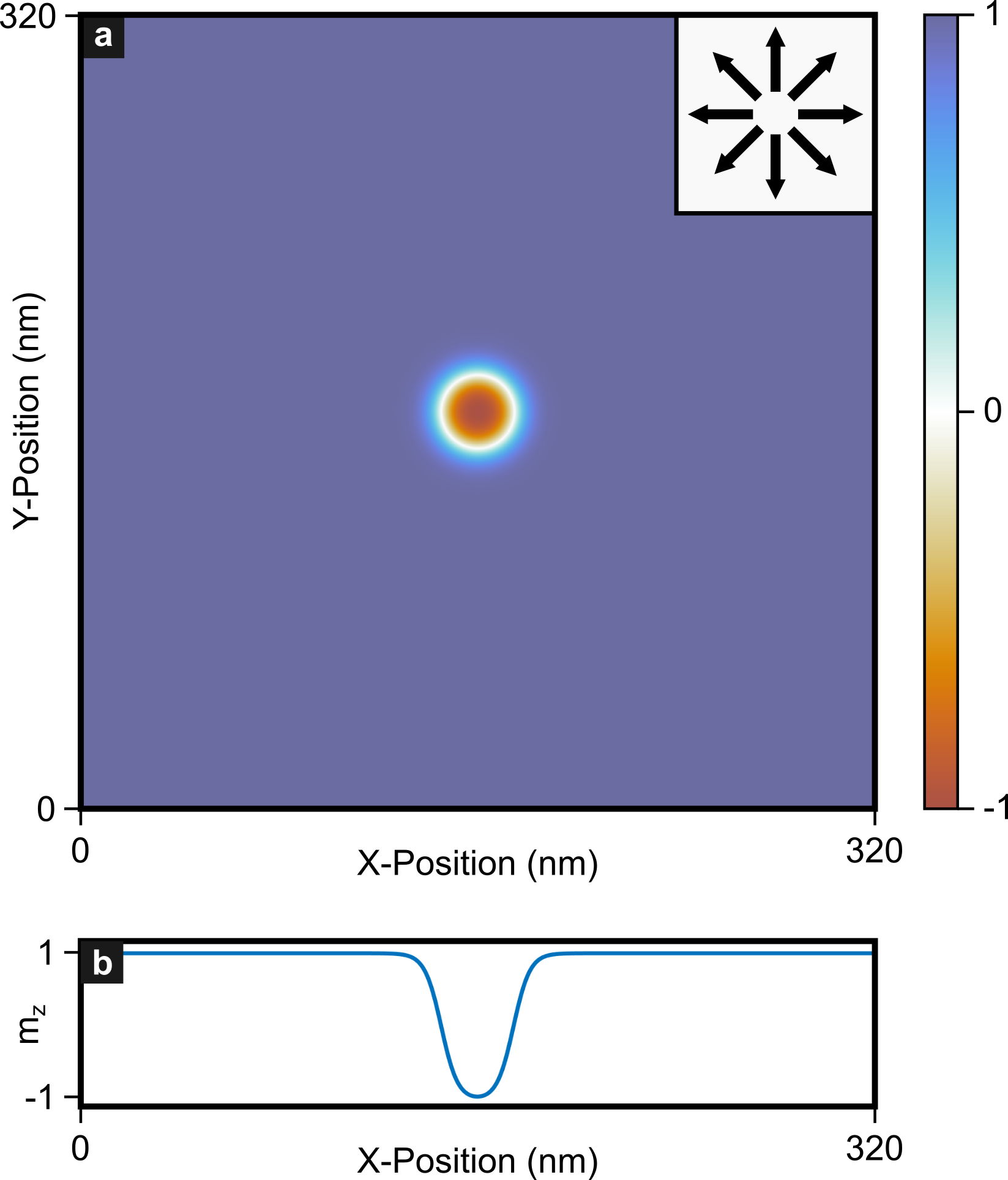}
\caption{\label{fig:ansatz}(a) Out-of-plane magnetization component $m_\mathrm{z}$ rendered as a color map for a $320~\mathrm{nm} \times 320~\mathrm{nm}$ thin film. A skyrmion is positioned in the center of the film according to the ansatz of Eq.~\eqref{eq:skyrmion_profile}. The inlay indicates the in-plane magnetization N\'{e}el-type configuration for the skyrmion. (b) One-dimensional cross-section of $m_\mathrm{z}$ taken through the skyrmion center and parallel to the $x$-axis.}.
\end{figure}

We simulate the annihilation of an isolated Néel-type skyrmion in a 2D magnetic material confined to the $xy$-plane. The initial skyrmion configuration is defined using the ansatz described in Refs.~\cite{Wang2018, Camley2023DMI}. The magnetization profile is isotropic and described by the radial function
\begin{eqnarray}
    \label{eq:skyrmion_profile}
    \theta(r) &=& 2 \arctan\left( \exp\left( \frac{r - R}{\Delta} \right) \right)\nonumber\\&&+ 2 \arctan\left( \exp\left( \frac{r + R}{\Delta} \right) \right) + \frac{(\mathcal{P} + 1)\pi}{2},
\end{eqnarray}
where $R$ is the skyrmion radius and $\Delta$ describes its domain-wall width. The angle is measured from the positive $z$-direction and $\mathcal{P} = \pm 1$ determines whether the core of the skyrmion points in the positive or negative $z$-direction. All skyrmions simulated here have $\mathcal{P} = -1$, so the last term of Eq.~\eqref{eq:skyrmion_profile} vanishes, thus the skyrmion's profile varies smoothly from $m_z = -1$ at its center to $m_z = +1$ at the edge of the simulation domain.

The skyrmion radius is given by
\begin{equation}
    \label{eq:skyrmion_radius}
    R = \pi |D|\sqrt{\frac{A}{16 A K^2_\mathrm{eff} - \pi^2 D^2 K_\mathrm{eff}}},
\end{equation}
with the effective anisotropy term $K_\mathrm{eff}$ given in terms of the uniaxial anisotropy field $H_\mathrm{k}$
\begin{equation}
    \label{eq:k_effective}
    K_\mathrm{eff} = \frac{1}{2} \mu_0 M_s H_k - \frac{1}{2} \mu_0 M_s^2.
\end{equation}

Finally, the characterization of the domain-wall width is given by
\begin{equation}
    \label{eq:domain_width}
    \Delta = \frac{\pi D}{4 K_\mathrm{eff}}.
\end{equation}

The material parameters used here are identical to those used in the dispersion relation simulations (Fig.~\ref{fig:dispersions}). However, the simulation domain is now a square grid of size $320~\mathrm{nm} \times 320~\mathrm{nm}$, with the skyrmion's core positioned on one of the four cells at the center of the grid. In other words, we ensure that $m_z=-1$ in one cell. This overall size of the material is maintained regardless of the cell size.

Figure~\ref{fig:ansatz}(a) shows an out-of-plane view of the skyrmion configuration, visualized as a color map of the $m_\mathrm{z}$ component of the magnetization. A corresponding cross-section of the $m_z$ magnetization component through the skyrmion's center is provided in Fig.~\ref{fig:ansatz}(b), highlighting the steep gradient at the domain wall. This state provides the common starting point for the subsequent investigation of field-driven contraction and annihilation dynamics across all models and discretizations.

\begin{figure}
\includegraphics[width=3.3in]{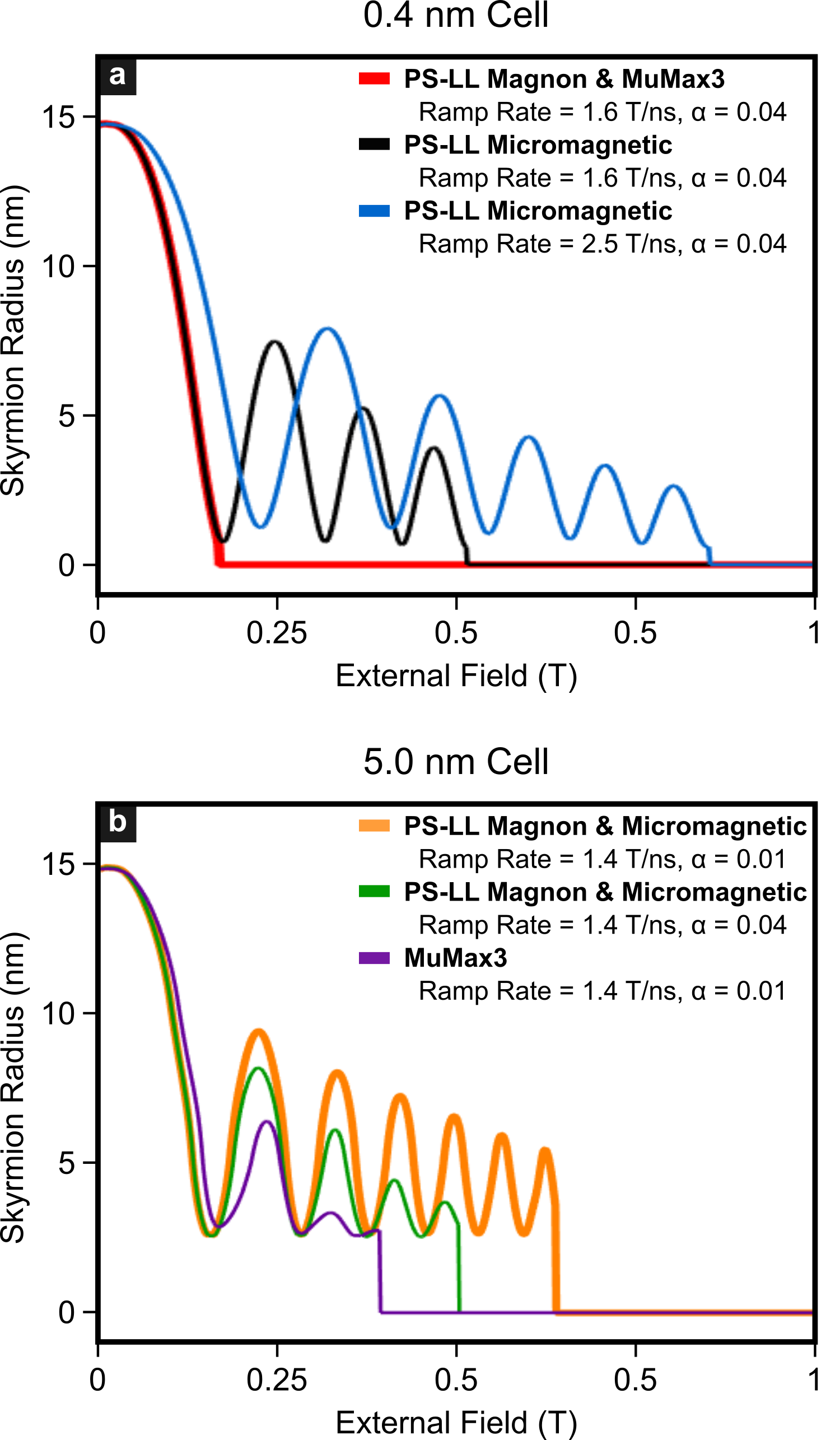}
\caption{\label{fig:radius_v_field} (a)~Skyrmion radius versus external field for a $0.4$ nm cell size. The red curve uses the PS-LL model with the magnon kernel and MuMax3. The black curve is the PS-LL model with the micromagnetic kernel with the same ramp rate and alpha as the red curve. The blue curve is also the micromagnetic kernel but with a larger ramp rate. (b)~Skyrmion radius versus external field for a $5.0$ nm cell size. Both PS-LL models behave nearly identically at a $5.0$ nm cell size. The orange and green curves individually represent both PS-LL models but for different values for the Gilbert damping constant $\alpha$. The purple curve is for MuMax3 with the same parameters as the orange curve. }
\end{figure}

{To investigate the dynamic annihilation of a single Néel skyrmion numerically, we use a linearly increasing applied magnetic field.} %Building on the initial configuration described above, we now investigate the dynamic response of a single Néel skyrmion to an applied magnetic field.
At $t = 0$ of each simulation, the system is initialized in a skyrmion state with no external field present. Once the simulation begins, a uniform out-of-plane magnetic field is applied along the $+\hat{z}$ direction and increased linearly in time. This ramping field causes the skyrmion to contract and eventually leads to the skyrmion's annihilation. The evolution of the skyrmion and the required field for its annihilation are influenced by the numerical model, the spatial resolution of the simulation, the rate at which the field is increased, and the selected damping constant. {We note that this method is used to access dynamical regimes leading to skyrmion annihilation and to estimate the annihilation field. However, the precise annihilation field is better determined from simulations using a static field.}

{We begin our simulations with an atomistic cell size of $0.4$~nm and use a Gilbert damping constant of $\alpha=0.04$, consistent with the considered multilayers. The increasing out-of-plane magnetic field induces an initial contraction of the skyrmion. After reaching a critical size, the subsequent skyrmion dynamics depend on the underlying model. In particular,} the skyrmion either annihilates or enters a breathing regime, characterized by periodic oscillations in its radius.  Similar breathing behavior has been reported in Ref.~\cite{austrup1}, where the authors attribute the phenomenon to the coupling between the skyrmion's radius and helicity as the skyrmion continuously precesses through energetically varying configurations (transitioning between Bloch and Néel types). As the helicity rotates into energetically unfavorable states, the radius is forced to periodically expand and contract to minimize the total energy. We have likewise observed the skyrmion transition between Bloch and Néel types during its oscillations.

We quantify the dynamics by calculating both the skyrmion radius and its topological invariant, the skyrmion number $\mathcal{S}$. Due to the skyrmion being initialized with its core centered on a lattice site, the symmetry of the square lattice, and the isotropic nature of the Hamiltonian, there is no net translational torque on the skyrmion center. Consequently, the skyrmion's center of mass remains pinned to the grid point throughout the shrinking dynamics. 

For the 0.4 nm cell size simulations, the skyrmion radius is calculated via its topological charge density~\cite{Braun2012}
\begin{equation}
    \label{eq:topdens}
    n = \iint\mathbf{m}\cdot\left(\partial_x\mathbf{m}\times\partial_y\mathbf{m}\right)dxdy.
\end{equation}
The resulting charge density peaks at the skyrmion's waist. We then perform an azimuthal average using a narrow Gaussian ring centered on the skyrmion core. This procedure is similar to that used to analyze x-ray scattering, e.g.,~\cite{ZhouHagstrom2022b,Jangid2023}, and results in a plot of topological charge density versus distance. The radius is determined by fitting a Gaussian function of the form
\begin{equation}
    \label{eq:Gaussian}
    f(r) = A e^{-(r-r_0)^2/(2\sigma^2)},
\end{equation}
{where $A$ is an amplitude, $r_0$ is the skyrmion's radius, and $\sigma$ is the skyrmion's width. The error associated with each quantity is found directly from the nonlinear least square minimization.}

For the 5.0 nm cell size simulations, the Gaussian fit method used before proved to be inaccurate due to the large discretization. For these simulations, we calculated the radius by examining the cross-section of the domain that passed through the skyrmion's center and was parallel to the x-axis. We found the cells where the out-of-plane magnetization transitions from positive to negative and performed a linear interpolation to find the radius as the distance from center of the skyrmion where the magnetization lied in-plane.

The evolution of the skyrmion radius is accompanied by the calculation of the topological charge to determine its annihilation unambiguously. The net topological charge can be calculated from Eq.~\eqref{eq:topdens}. However, this calculation is known to give rise to non-integer numbers when evaluated on a finite grid~\cite{Kim2020}. To avoid this issue, we compute the topological charge using an inverse stereographic projection of the material's magnetization onto the surface of a unit sphere~\cite{Kim2020}. In this approach, the 2D magnetic lattice is mapped onto the unit sphere, centered at the origin, with the material positioned in the first quadrant to simplify the calculation of azimuthal angles. Each square cell of the grid is split into two triangles for surface triangulation. To close the sphere, a virtual point is added at the north pole, representing the magnetization infinitely far from the skyrmion (assumed to be fully out-of-plane). Additional triangles are formed between this pole and each boundary edge segment. The solid angle subtended by each triangle is evaluated using~\cite{Kim2020,Rana2023}
\begin{equation}
    \label{eq:solidAngle}
    \Omega = 2 \arctan \left( \frac{ \mathbf{m}_1 \cdot (\mathbf{m}_2 \times \mathbf{m}_3) }{1 + \mathbf{m}_1 \cdot \mathbf{m}_2 + \mathbf{m}_2 \cdot \mathbf{m}_3 + \mathbf{m}_3 \cdot \mathbf{m}_1} \right)
\end{equation}
Leading to the topological charge
\begin{equation}
    \label{eq:topcharge}
    \mathcal{S} = \frac{1}{4\pi} \sum_{\text{triangles}} \Omega
\end{equation}

Figure~\ref{fig:radius_v_field}(a) shows several representative examples of the dynamic evolution of the skyrmion radius due to the increasing applied magnetic field and using atomic spatial resolution. In all cases, the skyrmion radius steadily decreases during the initial contraction phase, reaching a critical size of approximately $0.92$~nm on average, which is equivalent to $2.3$ cells. Beyond this point, the models diverge in their behavior. 

For both the PS-LL model with the magnon kernel and MuMax3 at atomic resolution, the skyrmion annihilates immediately after the initial contraction. {Annihilation is defined by a topological charge transition from 1 to 0, accompanied by energy release in the form of outward propagating magnons. This immediate annihilation occurs because the Zeeman energy surpasses the available exchange energy in these models.} 

\begin{figure*}[t]
    \centering
    \includegraphics[width=\linewidth]{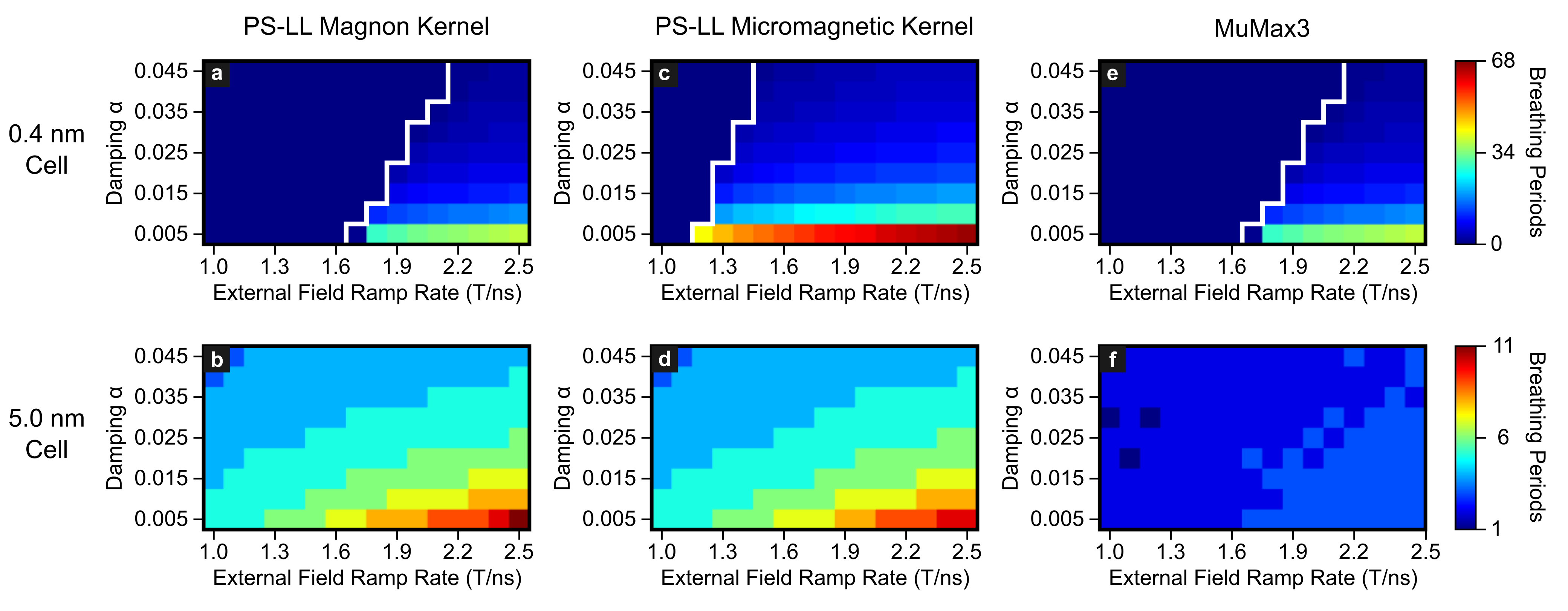}
    \caption{\label{fig:bounce} Color-plots indicated the number of breathing periods the skyrmion undergoes prior to annihilation. The top row pertains to simulations using a discretization of 0.4 nm for (a) magnon kernel, (c) micromagnetic kernel, and (e) MuMax3. The dark blue region to the left of the white line corresponds to zero breathing periods, signifying immediate annihilation following the initial contraction. The region to the right of the dividing line shows how the number of breathing periods increases with increasing ramp rate and decreasing damping. The bottom row has a discretization of 5.0 nm for (b) magnon kernel, (d) micromagnetic kernel, and (f) MuMax3. In this case, breathing is observed across the entire parameter space, with the number of periods generally increasing with higher ramp rates and lower damping.}
\end{figure*}

In contrast, the PS-LL model with the micromagnetic kernel preserves the skyrmion topology beyond this critical point. {In this case, the available exchange energy is unphysically large due to the $k^2$ approximation of the dispersion relation, which is invalid at atomic resolution}. Instead of annihilation, the skyrmion enters a breathing regime before {eventually annihilating at a much higher external field magnitude. The number and amplitude of these breathing oscillations depend on the applied field ramp rate. For the cases shown in Fig.~\ref{fig:radius_v_field}(a), increasing the ramp rate from 1.6~T/ns (black line) to 2.5~T/ns (blue line) increases the number of oscillations from 3 to 5 respectively. A more comprehensive view of how ramp rate and damping affect the number of breathing oscillations is shown in Fig.~\ref{fig:bounce}.} 

Representative simulations at a coarser resolution of 5~nm are shown in Fig.~\ref{fig:radius_v_field}(b). At this cell size, all simulations exhibit skyrmion breathing prior to annihilation. The two PS-LL models are effectively indistinguishable, whereas MuMax3 is quantitatively different. The agreement between the two PS-LL models at larger cell sizes arises because the range of allowed wavevectors is reduced and confined to a small section of the first Brillouin zone where the micromagnetic approximation is largely accurate. The quantitative disagreement of the simulations performed by MuMax3 can be attributed to the cell size being above the exchange length. However, we note that the simulation was numerically stable in this case. Therefore, the skyrmion dynamics presented here are accurate within the exchange interaction model and its concomitant energy.

\begin{figure*}[t]
\includegraphics[width=\linewidth]{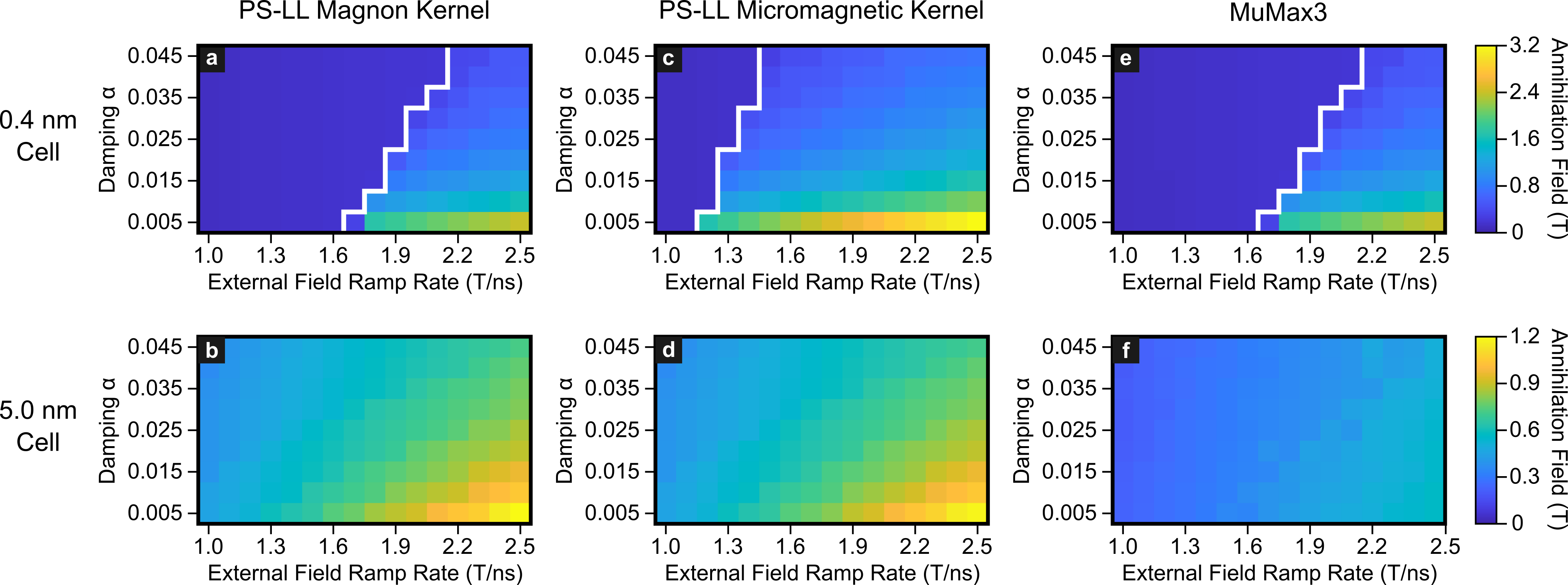}
\caption{\label{fig:phase_plots} Color-plots indicating the field at which the skyrmion annihilates based on a change of the topological charge from 1 to 0. The top row pertains to simulations using a discretization of 0.4~nm for (a) magnon kernel, (c) micromagnetic kernel, and (e) MuMax3. The white boundary indicates the transition between skyrmion annihilation and breathing. To the left of the dividing line, the skyrmion annihilates at a relatively low field. To the right of the dividing line, the skyrmion's annihilation field increases significantly with increasing ramp rate and decreasing damping. The bottom row has a discretization of 5.0~nm for (b) magnon kernel, (d) micromagnetic kernel, and (f) MuMax3. In this case, we observe annihilation only after few breathing periods but the annihilation field shows the same trend with respect to ramp rate and damping.}
\end{figure*}

\section{Phase diagram of skyrmion annihilation}

We now examine how the {dynamic skyrmion annihilation} is influenced by the external field ramp rate and the Gilbert damping constant $\alpha$. As discussed earlier, all skyrmions eventually annihilate under a sufficiently strong out-of-plane magnetic field. However, the pathway towards annihilation can vary significantly: in some cases, the skyrmion collapses directly following its initial contraction, while in others it undergoes {one or multiple breathing periods} before ultimately annihilating. These distinct behaviors are mapped out in the phase diagrams of Fig.~\ref{fig:phase_plots} and Fig.~\ref{fig:bounce}. For both figures, results are shown for all three models, PS-LL with the magnon kernel, PS-LL with the micromagnetic kernel, and MuMax3, at both atomic-scale ($0.4$~nm) and micromagnetic-scale ($5.0$~nm) discretizations.

The color in Fig.~\ref{fig:phase_plots} indicates the external field strength at which annihilation occurs, plotted as a function of ramp rate and damping. A notable feature of the $0.4$~nm phase diagrams is the near-identical behavior of the PS-LL magnon kernel, Fig.~\ref{fig:phase_plots}(a), and MuMax3, Fig.~\ref{fig:phase_plots}(e). This agreement is directly attributable to their nearly identical dispersion relations, c.f. Fig.~\ref{fig:dispersions}(a) and Fig.~\ref{fig:dispersions}(e) at this discretization. In contrast, the PS-LL micromagnetic kernel, Fig.~\ref{fig:dispersions}(c) shows noticeably different behavior at the same cell size.

In the $0.4$ nm phase plots, Figs.~\ref{fig:dispersions}(a), (c), and (e), the parameter space is clearly partitioned into two regions, for which we have indicated the boundary with a white line. On the left side of this line, skyrmions annihilate immediately after their initial contraction. On the right, they undergo multiple breathing events before annihilation. The exact number of breathing periods is quantified in Figs.~\ref{fig:bounce}(a), (c), and (e) respectively. We observe that the number of oscillations increases with higher ramp rates and lower damping.

The boundary for the PS-LL micromagnetic kernel, Fig.~\ref{fig:dispersions}(c), is shifted leftward compared to the other two models, indicating that the breathing region begins at lower ramp rates. This shift is related to the differences in the dispersion relations of the different models. At the 0.4~nm cell size, the dispersion relation for the micromagnetic kernel deviates from the other two models as the curve approaches the edges of the Brillouin zone. This deviation shows that the micromagnetic approximation fails to accurately represent short-wavelength magnons. As a result, the PS-LL micromagnetic model overestimates the energy cost of magnons at high $k$, effectively increasing the stiffness of the system and inhibiting immediate collapse. In other words, the exchange energy of a skyrmion approaching a singularity is unphysically large. As a result, the exchange energy is released, much like a spring, causing the skyrmion to breathe instead of annihilate. This also explains why the fields required for annihilation are generally larger for the micromagnetic kernel compared to the other two models as this deference to breathing occurs each time the skyrmion reaches its critical size. 

At the larger cell size of $5.0$~nm, Figs.~\ref{fig:phase_plots}(b), (d) and (f), the two PS-LL kernels are nearly indistinguishable, while MuMax3 diverges. This again mirrors the dispersion relations observed in Figs.~\ref{fig:dispersions}(b), (d) and (f), where the PS-LL kernels closely match one another but differ from MuMax3. At this discretization, the micromagnetic approximation becomes more appropriate, and both PS-LL kernels produce nearly parabolic dispersions. Interestingly, the phase diagrams at this discretization are not divided into two regions like their atomic scale counterparts: skyrmion breathing is always observed, as shown in Figs.~\ref{fig:bounce}(b), (d), and (f) respectively. The lack of distinct regions occurs because a larger cell size assumes that atomic spins are uniformly aligned within each cell, reducing the exchange energy contribution. Therefore, as the skyrmion contracts, a singularity is established at its core. This is seen in Fig.~\ref{fig:radius_v_field}(b) from the minimum skyrmion radius of $2.5$~nm (half a cell) that results from our calculation method when two adjacent cells have antiparallel spins. Because of its topological barrier, the external field is not able to switch the singularity and breathing ensues. Once the external field is sufficient to switch the singularity, the skyrmion annihilates. For this reason, we observe that the skyrmions tend to annihilate at lower field amplitudes, when a larger cell size is used. In simulations performed by MuMax3, annihilation is observed at fields approximately a factor of two smaller than what was seen at a $0.4$~nm cell size. According to our interpretation, this is expected, as the cosine-like dispersion for a $5.0$ nm cell size underestimates the exchange energy at the smallest resolved wavevectors.

Further evidence highlighting the importance of the exchange energy description is found in the general behavior of the annihilation fields as a function of damping. In all cases, the highest annihilation fields occur for high field ramp rates and low damping values. On the one hand, the ramp rates used here are comparable to the characteristic timescales of the dynamics themselves, causing the magnetic system to lag in its response. On the other hand, low damping extends the lifetime of fast dynamics. Contrary to naive expectations, this implies that a nonlinear structure like a skyrmion requires more time to stabilize or, in this context, to annihilate. Since the damping is higher, the high-frequency excitations are more rapidly quenched allowing the skyrmion to reach its new equilibrium state more quickly. This interpretation holds insofar as damping is sufficiently low for the explicit Landau-Lifshitz equation to remain valid.

\section{Conclusion}

Our investigation demonstrates that the choice of numerical model and spatial discretization has a significant impact on skyrmion annihilation pathways and the physical limits of topological protection. {Reducing the spatial discretization is generally beneficial, as also demonstrated in recent investigations using multiscale approaches~\cite{DeLucia2017,Winkler2021} showing convergence between micromagnetic and atomistic or Heisenberg models~\cite{Winkler2025}. Our} results also showcase the importance of how the exchange interaction is implemented. We examined MuMax3, a widely used micromagnetic solver. {The micromagnetic approach is valid when textures are longer than the exchange length, but numerical implementations typically require a cell-size smaller than the exchange length to accurately solve the Laplacian operator. In the particular case of MuMax3,} the exchange implementation leads to a cosine-type dispersion relation, which misrepresents the exchange energy landscape. Interestingly, this implementation turns out to be correct at atomic resolution insofar as only nearest-neighbor interactions are required. More generally, next-nearest-neighbor interactions have been shown to be relevant from density functional theory~\cite{Eriksson2017}. {In such cases, the PS-LL model is an attractive solution because it is grid-independent and captures the necessary physics in its convolution kernel. For example, in this work, we extended the previous 1D implementation~\cite{rockwell} to 2D and included the DMI from first principles calculations. It is also possible to consider other nonlocal interactions such as the magnetostatic approximation for spin waves~\cite{Roxburgh2025}. Therefore, other nonlocal interactions, such as a next-nearest-neighbor exchange, can be represented in Fourier space.}

Our analysis focused primarily on the effect of exchange energy. However, other physical contributions not considered here are important for accurate physical models, e.g., nonlocal dipole fields and temperature. Nonlocal dipolar fields lead to field closures which can reinforce the skyrmion as it contracts. This mechanism is exploited, for example, to stabilize skyrmions in materials lacking DMI but with sufficient thickness~\cite{Montoya2017}. A comprehensive analysis would require a full three-dimensional pseudospectral approach. Temperature can play an important role, as fluctuations may perturb the magnetization at the skyrmion core and lead to its premature annihilation. In such cases, a statistical analysis would need to be implemented to define a mean skyrmion lifetime.

A fundamental question addressed in this work is the relationship between small-amplitude waves and nonlinear structures. Specifically, the dispersion relation is particular of small-amplitude waves in the sense that the excitation is linear, so that there is no spectral mixing. Regardless, {the dispersion relation} reflects the structure of the kernel. This kernel then applies to nonlinear structures where mixing is allowed. This is the fundamental reason why the details of the dispersion relation past half the FBZ are relevant for nonlinear problems and why skyrmions are better resolved using the PS-LL method. In other words, the Fourier representation of the kernel comprises all available spatial frequencies in the physical system, which are necessary to define a solitary wave. We expect this principle to extend to other textures in 2D and 3D, as well as many other physical systems where multiple scales are ubiquitous and difficult to encapsulate in a single model or approximation. Therefore, this study is a stepping stone towards future progress to model, understand, and manipulate topological textures in a physically accurate manner. From a broader perspective, the accurate modeling of the energy and momentum space will lead to improved multiscale models, including ultrafast dynamics, while the model of nonlocal interaction will enable further developments in material discovery using machine learning approaches.

\section*{Acknowledgments}

This work was supported by the U.S. Department of Energy, Office of Basic Energy Sciences under Award Number DE-SC0024339.

\section*{Data Availability}

The data that support the findings of this article are openly available~\cite{OSF}.

\appendix
\section{Derivation of the 2D Dzyaloshinskii-Moriya Interaction PS-LL Kernel}
\label{sec:DMI}

\begin{figure}[b]
    \centering
    \includegraphics[width=0.8\linewidth]{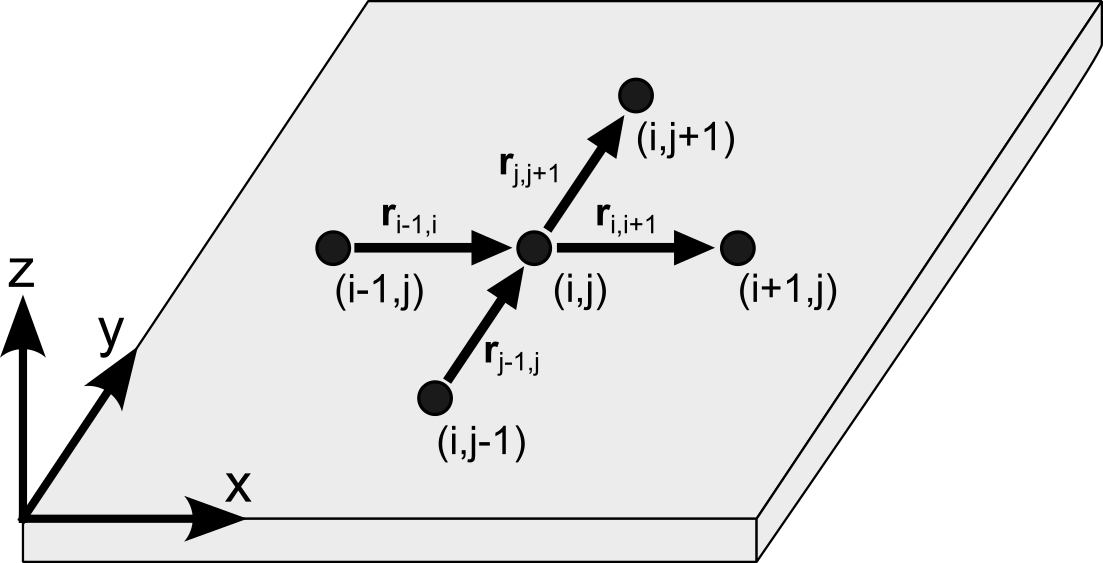}
    \caption{\label{fig:Appendix} Geometry to derive the DMI based on an element at position $\{i,j\}$. The distances are represented by the vector \textbf{r}.}
\end{figure}
Consider a 2D ferromagnet in the $\hat{x}\hat{z}$-plane that is subject to an interfacial Dzyaloshinskii–Moriya interaction (DMI) with the symmetry breaking occurring in the $\hat{z}$ direction. The DMI energy for a site $\mathbf{S}_{(i,j)}$ due to its four nearest neighbor interactions is given by
\begin{eqnarray}
U_{DMI} &=& \mathbf{D}_{(i-1,i)} \cdot \left(\mathbf{S}_{(i-1,j)}\times\mathbf{S}_{(i,j)}\right) + \nonumber\\ 
&&\mathbf{D}_{(i,i+1)} \cdot\left(\mathbf{S}_{(i,j)}\times\mathbf{S}_{(i+1,j)}\right) + \nonumber\\ 
&&\mathbf{D}_{(j-1,j)} \cdot \left(\mathbf{S}_{(i,j-1)}\times\mathbf{S}_{(i,j)}\right) + \nonumber\\ 
&&\mathbf{D}_{(j,j+1)} \cdot\left(\mathbf{S}_{(i,j)}\times\mathbf{S}_{(i,j +1)}\right)
\end{eqnarray}

Performing the cross products
\begin{subequations}
\begin{eqnarray}
\mathbf{S}_{(i-1,j)}\times\mathbf{S}_{(i,j)} &=&
\begin{vmatrix}
\hat{x} & \hat{y} & \hat{z} \\ 
S_{x(i-1,j)} & S_{y(i-1,j)} & S_{z(i-1,j)} \\ 
S_{x(i,j)} & S_{y(i,j)} & S_{z(i,j)}
\end{vmatrix} = \\
&&\left(S_{y(i-1,j)}S_{z(i,j)} - S_{z(i-1,j)}S_{y(i,j)}\right)\hat{x}+ \nonumber\\
&&\left(S_{z(i-1,j)}S_{x(i,j)} - S_{x(i-1,j)}S_{z(i,j)}\right)\hat{y}+ \nonumber\\
&&\left(S_{x(i-1,j)}S_{y(i,j)} - S_{y(i-1,j)}S_{x(i,j)}\right)\hat{z},\nonumber
\end{eqnarray}

\begin{eqnarray}
\mathbf{S}_{(i,j)}\times\mathbf{S}_{(i+1,j)} &=& 
\begin{vmatrix}
\hat{x} & \hat{y} & \hat{z} \\ 
S_{x(i,j)} & S_{y(i,j)} & S_{z(i,j)} \\ 
S_{x(i+1,j)} & S_{y(i+1,j)} & S_{z(i+1,j)}
\end{vmatrix} = \\
&&\left(S_{y(i,j)}S_{z(i+1,j)} - S_{z(i,j)}S_{y(i+1,j)}\right)\hat{x}+ \nonumber\\
&&\left(S_{z(i,j)}S_{x(i+1,j)} - S_{x(i,j)}S_{z(i+1,j)}\right)\hat{y}+ \nonumber\\
&&\left(S_{x(i,j)}S_{y(i+1,j)} - S_{y(i,j)}S_{x(i+1,j)}\right)\hat{z},\nonumber 
\end{eqnarray}

\begin{eqnarray}
\mathbf{S}_{(i,j-1)}\times\mathbf{S}_{(i,j)} &=&
\begin{vmatrix}
\hat{x} & \hat{y} & \hat{z} \\ 
S_{x(i,j-1)} & S_{y(i,j-1)} & S_{z(i,j-1)} \\ 
S_{x(i,j)} & S_{y(i,j)} & S_{z(i,j)}
\end{vmatrix} = \\
&&\left(S_{y(i,j-1)}S_{z(i,j)} - S_{z(i,j-1)}S_{y(i,j)}\right)\hat{x}+\nonumber \\
&&\left(S_{z(i,j-1)}S_{x(i,j)} - S_{x(i,j-1)}S_{z(i,j)}\right)\hat{y}+\nonumber \\
&&\left(S_{x(i,j-1)}S_{y(i,j)} - S_{y(i,j-1)}S_{x(i,j)}\right)\hat{z},\nonumber 
\end{eqnarray}

\begin{eqnarray}
\mathbf{S}_{(i,j)}\times\mathbf{S}_{(i,j+1)} &=& 
\begin{vmatrix}
\hat{x} & \hat{y} & \hat{z} \\ 
S_{x(i,j)} & S_{y(i,j)} & S_{z(i,j)} \\ 
S_{x(i,j+1)} & S_{y(i,j+1)} & S_{z(i,j+1)}
\end{vmatrix} = \\
&&\left(S_{y(i,j)}S_{z(i,j+1)} - S_{z(i,j)}S_{y(i,j+1)}\right)\hat{x}+\nonumber \\
&&\left(S_{z(i,j)}S_{x(i,j+1)} - S_{x(i,j)}S_{z(i,j+1)}\right)\hat{y}+\nonumber \\
&&\left(S_{x(i,j)}S_{y(i,j+1)} - S_{y(i,j)}S_{x(i,j+1)}\right)\hat{z}.\nonumber 
\end{eqnarray}
\end{subequations}

The DMI vectors are given by the cross product between the direction of the vector $\mathbf{r}$ connecting two points (shown in Fig.~\ref{fig:Appendix}) and the symmetry breaking direction
\begin{subequations}
\begin{eqnarray}
\mathbf{D}_{(i-1,i)} &=& \mathbf{D}_{(i,i+1)} = D(\hat{r}_{(i-1,i)} \times \hat{z}) \nonumber\\ 
&=& D(\hat{x} \times \hat{z}) = -D\hat{y}, \\
\mathbf{D}_{(j-1,j)} &=& \mathbf{D}_{(j,j+1)} = D(\hat{r}_{(j-1,j)} \times \hat{z}) \nonumber\\ 
&=& D(\hat{y} \times \hat{z}) = D\hat{x}.
\end{eqnarray}
\end{subequations}

After performing the dot product, we obtain the DMI energy
\begin{eqnarray}
U_{DMI} &=& D\big(-S_{z(i-1,j)}S_{x(i,j)} + S_{x(i-1,j)}S_{z(i,j)}\nonumber\\ 
&& - S_{z(i,j)}S_{x(i+1,j)} + S_{x(i,j)}S_{z(i+1,j)}\nonumber\\ 
&& +S_{y(i,j-1)}S_{z(i,j)} - S_{z(i,j-1)}S_{y(i,j)}\nonumber\\ 
&& + S_{y(i,j)}S_{z(i,j+1)} - S_{z(i,j)}S_{y(i,j+1)}\big)\nonumber\\
&=& D\big[S_{x(i,j)}(S_{z(i+1,j)}-S_{z(i-1,j)})\nonumber\\ 
&& + S_{y(i,j)}(S_{z(i,j+1)} - S_{z(i,j-1)})\nonumber\\ 
&& + S_{z(i,j)}(S_{x(i-1,j)} - S_{x(i+1,j)}\nonumber\\ 
&& + S_{y(i,j-1)} - S_{y(i,j+1)})].
\end{eqnarray}

The effective field from DMI acting on site $\mathbf{S}_{(i,j)}$ is given by
\begin{equation}
\mathbf{H}_{DMI} = -\frac{\partial U_{DMI}}{\partial S_{x(i,j)}}\hat{x} - \frac{\partial U_{DMI}}{\partial S_{y(i,j)}}\hat{y} - \frac{\partial U_{DMI}}{\partial S_{z(i,j)}}\hat{z}.
\end{equation}

After taking the derivatives, we get the effective field acting on site $\mathbf{S}_{(i,j)}$ due to DMI from its nearest neighbors
\begin{subequations}
\begin{eqnarray}
H_{x} &=& D\left(S_{z(i-1,j)} - S_{z(i+1,j)}\right), \\ 
H_{y} &=& D\left(S_{z(i,j-1)} - S_{z(i,j+1)}\right), \\ 
H_{z} &=& D\Big(S_{x(i+1,j)} - S_{x(i-1,j)}\nonumber\\ 
&& + S_{y(i,j+1)} - S_{y(i,j-1)}\Big).
\end{eqnarray}
\end{subequations}

We now make the assumption that site $\mathbf{S}_{(i,j)}$ undergoes wave motion and that the neighboring sites experience the same motion with a phase shift
\begin{subequations}
\begin{eqnarray}
S_{z(i-1,j)} &=& S_{z(i,j)}e^{-i(k_x a)}, \\
S_{z(i+1,j)} &=& S_{z(i,j)}e^{i(k_x a)}, \\
S_{z(i,j-1)} &=& S_{z(i,j)}e^{-i(k_y a)}, \\
S_{z(i,j+1)} &=& S_{z(i,j)}e^{i(k_y a)}, \\
S_{x(i+1,j)} &=& S_{x(i,j)}e^{i(k_x a)}, \\
S_{x(i-1,j)} &=& S_{x(i,j)}e^{-i(k_x a)}, \\
S_{y(i,j+1)} &=& S_{y(i,j)}e^{i(k_y a)}, \\
S_{y(i,j-1)} &=& S_{y(i,j)}e^{-i(k_y a)}.
\end{eqnarray}
\end{subequations}

Where $a$ is the spacing between atoms. Putting these into the effective field terms gives
\begin{subequations}
\begin{eqnarray}
H_{x} &=& D\left(S_{z(i,j)}e^{-i(k_x a)} - S_{z(i,j)}e^{i(k_x a)}\right), \\ 
H_{y} &=& D\left(S_{z(i,j)}e^{-i(k_y a)} - S_{z(i,j)}e^{i(k_y a)}\right), \\ 
H_{z} &=& D\Big(S_{x(i,j)}e^{i(k_x a)} - S_{x(i,j)}e^{-i(k_x a)}\nonumber\\ 
&& + S_{y(i,j)}e^{i(k_y a)} - S_{y(i,j)}e^{-i(k_y a)}\Big).
\end{eqnarray}
\end{subequations}

After trigonometric substitutions, we get
\begin{subequations}
\begin{eqnarray}
H_{x} &=& -S_{z(i,j)}D2i\sin{(k_x a)}, \\ 
H_{y} &=& -S_{z(i,j)}D2i\sin{(k_y a)}, \\ 
H_{z} &=& S_{x(i,j)}D2i\sin{(k_x a)}\nonumber\\ 
&& + S_{y(i,j)}D2i\sin{(k_y a)}. 
\end{eqnarray}
\end{subequations}

The DMI contributions to the PS-LL kernel become
\begin{subequations}
\begin{eqnarray}
dx:&\quad& \mathcal{F}^{-1}\left[-\hat{S}_{z(i,j)}D2i\sin{(k_x a)}\right],\\ 
dy:&\quad& \mathcal{F}^{-1}\left[-\hat{S}_{z(i,j)}D2i\sin{(k_y a)}\right], \\ 
dz:&\quad& \mathcal{F}^{-1}\Big[\hat{S}_{x(i,j)}D2i\sin{(k_x a)}\nonumber\\ 
&& + \hat{S}_{y(i,j)}D2i\sin{(k_y a)}\Big].
\end{eqnarray}
\end{subequations}

\end{document}